\documentclass[prl,twocolumn]{revtex4-1}

\usepackage[margin=1in]{geometry}                
\usepackage{amsmath}
\usepackage{amssymb}
\usepackage{graphicx}
\usepackage{setspace}
\usepackage{xcolor}

\newcommand{\xor}{\oplus}

\newenvironment{not-sure}
  {\newline\textbf{Not Sure}\newline\it}
  {\rm\newline\textbf{End Not Sure}}

\newcommand{\be}{\begin{equation}}

\newcommand{\ee}{\end{equation}}
\newcommand{\bes}{\begin{eqnarray}}
\newcommand{\ees}{\end{eqnarray}}

\newcommand{\figref}[1]{Figure~\ref{#1}}
\renewcommand{\eqref}[1]{Eq.~(\ref{#1})}

\newcommand{\avg}[1]{\left \langle #1 \right \rangle}

\begin{document}

\title{Logic and connectivity jointly determine criticality in biological gene regulatory networks}

\date{\today}                                           

\author{Bryan C. Daniels$^1$, Hyunju Kim$^{2,3}$, Douglas Moore$^3$, Siyu Zhou$^4$, Harrison Smith$^2$, Bradley Karas$^3$,  Stuart A. Kauffman$^5$, Sara I. Walker$^{1,2,3}$}
\affiliation{$^1$ASU--SFI Center for Biosocial Complex Systems, Arizona State University, Tempe, AZ 85287 \\$^2$ School of Earth and Space Exploration, Arizona State University, Tempe AZ 85287\\ $^3$ Beyond Center for Fundamental Concepts in Science, Arizona State University\\ $^4$ Department of Physics, Arizona State University, Tempe AZ 85287\\ $^5$ Institute for Systems Biology, Seattle WA}

\begin{abstract}

The complex dynamics of gene expression in living cells can be well-approximated
using Boolean networks. 
The average sensitivity is a natural measure of stability
in these systems: values below one indicate typically stable dynamics associated with an ordered phase, whereas values above one indicate chaotic dynamics. This yields a theoretically motivated adaptive advantage to being
near the critical value of one, at the boundary between order and chaos. Here, we measure average sensitivity for 66 publicly available 
Boolean network models describing the function of gene regulatory
circuits across diverse living processes.
We find the average sensitivity values for these networks are clustered around unity, indicating they are near critical. In many types of random networks, 
mean connectivity $\langle K \rangle $ and the average activity bias of the logic functions $\langle p \rangle$ have been found 
to be the most important network properties
in determining average sensitivity, and by extension a network's criticality. Surprisingly, many of 
these gene regulatory networks achieve the near-critical state with $\langle K \rangle $ and $\langle p \rangle$ far from that predicted for critical systems: randomized networks sharing the local causal structure and local logic of biological networks better reproduce their critical behavior than controlling for macroscale properties such as $\langle K \rangle $ and $\langle p \rangle$ alone.  This suggests the local properties of genes interacting within regulatory networks are selected to collectively be near-critical, and this non-local property of gene regulatory network dynamics cannot be predicted using the density of interactions alone.



\end{abstract}

\maketitle





One of the most widely discussed, and oft debated, aspects of the physics of life is the role of criticality \cite{adami1995self,munoz2017colloquium,krotov2014morphogenesis,hidalgo2014information,roli2018dynamical,bak2013nature}. Criticality, or tuning to a point of marginal stability, is hypothesized to drive both the robustness and evolvability of living processes \cite{Lan90,Kau93}.  Systems that are far from criticality are argued to be less adaptive, either being too stable to be responsive in the ordered phase, or being too unstable to maintain memory in the chaotic phase. Many biological systems are now known to be poised in between these regimes, with proximity to criticality reported across a variety of biological systems with very different functions, such as neural firing, animal motion and social behavior, and gene regulation \cite{DanKraFla17,beggs2008criticality,HalBeg05,mora2011biological,munoz2017colloquium,sole1999criticality,bialek2014social}. 
Despite the apparent ubiquity of criticality across many distinct living processes, precisely how the local properties of biological networks generate macroscale behavior that is collectively critical is not yet explained. 

Boolean networks are widely implemented dynamical systems for inferring the existence of critical behavior in biological systems.  In particular, the complex regulatory interactions describing the dynamics of genes essential to cellular function are well-approximated by Boolean models \cite{karlebach2008modelling,wang2012boolean,albert2008boolean,helikar2012cell,helikar2013cell,helikar2009chemchains,bornholdt2008boolean}. 
 Physiological parameters such as reaction rates are coarse-grained in Boolean models, meaning one does not need to know specific values that are in many cases intractable to experimentally measure. By reducing the number of fine-tuned parameters Boolean models are much simpler to build and simulate, yet they nonetheless capture many important dynamical features of real biological systems associated with their function \cite{davidich2008transition}. 
Boolean gene regulatory network models have also successfully predicted cellular behavior including the robustness of the cell-cycle, cell differentiation processes, and cellular response to DNA damage\cite{li2004yeast, davidich2008boolean, huang2005cell, choi2012attractor}.

Evidence biological gene regulatory networks operate near criticality has been so far limited to a handful of experimental examples, but these are increasing in frequency.  The effects of experimental perturbations of single genes in \emph{S. Cerevisiae} 
\cite{SerVilSem04,SerVilGra07}, the dynamics of gene expression in the macrophage \cite{NykPriAld08}, and a handful of networks with experimentally-derived network topology \cite{balleza2008critical,ChoLloSmo10} have been shown to be consistent with near-criticality.  These experimental cases provide isolated examples that do suggest criticality plays an important role in some gene networks. However, it is currently unknown how widespread criticality is across diverse gene regulatory networks with different structure and function. Addressing this question requires a systematic survey of different gene regulatory networks, as we provide here.

Before presenting our results, it is important to emphasize that observations of criticality in real systems have so far been primarily motivated by the theory of random Boolean networks (e.g. \cite{MorAma05,RamKesYli06,GouTeuGul12}). 
By constructing ensembles of random networks with fixed average in-degree $\langle K \rangle$ and average activity bias $\langle p \rangle$, one can readily determine thresholds for criticality as a function of $\langle K \rangle$ and $\langle p \rangle$ for the ensemble. The results indicate connectivity and the mean bias of Boolean logic functions both play a role in determining critical behavior. However, while the ensemble of random networks in these theoretical studies subsumes those we expect to exhibit biological function, the ensemble is not exclusive to living examples. Whether or not $\langle K \rangle$ and $\langle p \rangle$, as statistical characterizations of connectivity and logic, are specific enough to explain criticality in networks with biological function therefore remains to be tested. 


The recent proliferation of Boolean network models for functioning gene circuits now permits the possibility to directly address the drivers of criticality in real biological networks. In the current study, 66 Boolean models were obtained from the Cell Collective database \cite{cellcollective}.  The networks represent biological processes including virus and cell cycles, cell differentiation, cell plasticity, cell apoptosis, cell migration, and signalling pathways, among other gene regulatory functions. These genetic circuits encapsulate a wide range of fundamental biological processes across humans, animals, plants, bacteria and viruses, and range in size from 5 nodes to 321 nodes.

To infer criticality in these networks, we use a measure of average sensitivity \cite{ShmKau04}. 
We define the average sensitivity $s$ as in Ref.~\cite{ShmKau04},
starting with the Boolean derivative that measures the number
of inputs for which flipping a bit at timestep $t$ 
changes the value of the
output at timestep $t+1$, 
then averaging over nodes and over all possible
input states $\vec x$.  Defining the discrete dynamics as
$\vec x(t+1) = \vec f(\vec x(t))$, 
\begin{equation}
\label{seqn}
s = \frac{1}{N} \sum_i^N \avg{ \sum_j^N \frac{\partial f_i(\vec x)}{\partial x_j} }_{\vec x}.
\end{equation}
Here $\frac{\partial f_i(\vec x)}{\partial x_j}$ represents
the sensitivity of node $i$ to changing node $j$ when starting
in state $\vec x$ \cite{LuqSol00}:
\begin{equation}
\frac{\partial f_i(\vec x)}{\partial x_j} = f_i(x_1,...,x_j,... x_N) \xor f_i(x_1,...,\bar x_j,... x_N),
\end{equation}
with $\bar x_j$ representing the logical negation of $x_j$.
Defined in this way, the average sensitivity $s$ is the expected number of nodes changed at the next timestep given a perturbation that flips the state of one node at the current timestep.  It is equivalently equal to the average Hamming 
distance between the perturbed and unperturbed state at time $t+1$ 
when a random bit is flipped at time $t$ (see Supplemental Material).

The average sensitivity was defined in Ref.~\cite{ShmKau04} to be an indicator of the critical transition in random Boolean networks from an ordered to a chaotic phase. 
In an infinite ergodic system, this transition happens at $s=1$ \cite{VilCamDam16}\footnote{Finite size effects cause $s=1$ to correspond to a slightly sub-critical regime.}.
In the ordered phase, bit-flip perturbations have effects that become smaller over time, while in the chaotic phase these changes grow in time and spread to affect most of the network \cite{LuqSol00}.
The original results exploring this damage-spreading transition \cite{DerPom86,LuqSol00} 
make this connection analytically under two assumptions: (1) As in other spreading processes \cite{DanKraFla17}, 
this critical transition happens when the local measure of spreading (here, the average sensitivity) is equal to 1 only in the limit of $N \rightarrow \infty$, where finite-size effects of saturation are not important, and with the assumption of ergodicity \cite{VilCamDam16}; (2) the dynamics were assumed to be synchronous, with all nodes updated at each timestep.
We note that a large fraction of the models we test 
were not designed to be used with synchronous updating. Yet even in the asynchronous case, we expect $s=1$ to correspond to the damage spreading critical transition as $N \rightarrow \infty$.
In this limit and as $t \rightarrow \infty$, regardless of the specifics of how nodes are updated, we can treat the damage spreading as a simple branching process with branching ratio equal to the average sensitivity $s$.
Thus we expect that networks run asynchronously will have similar bulk behavior to those run synchronously, with a critical transition at $s = 1$ in the infinite limit.

\begin{figure}
\centering
\includegraphics[scale=0.375]{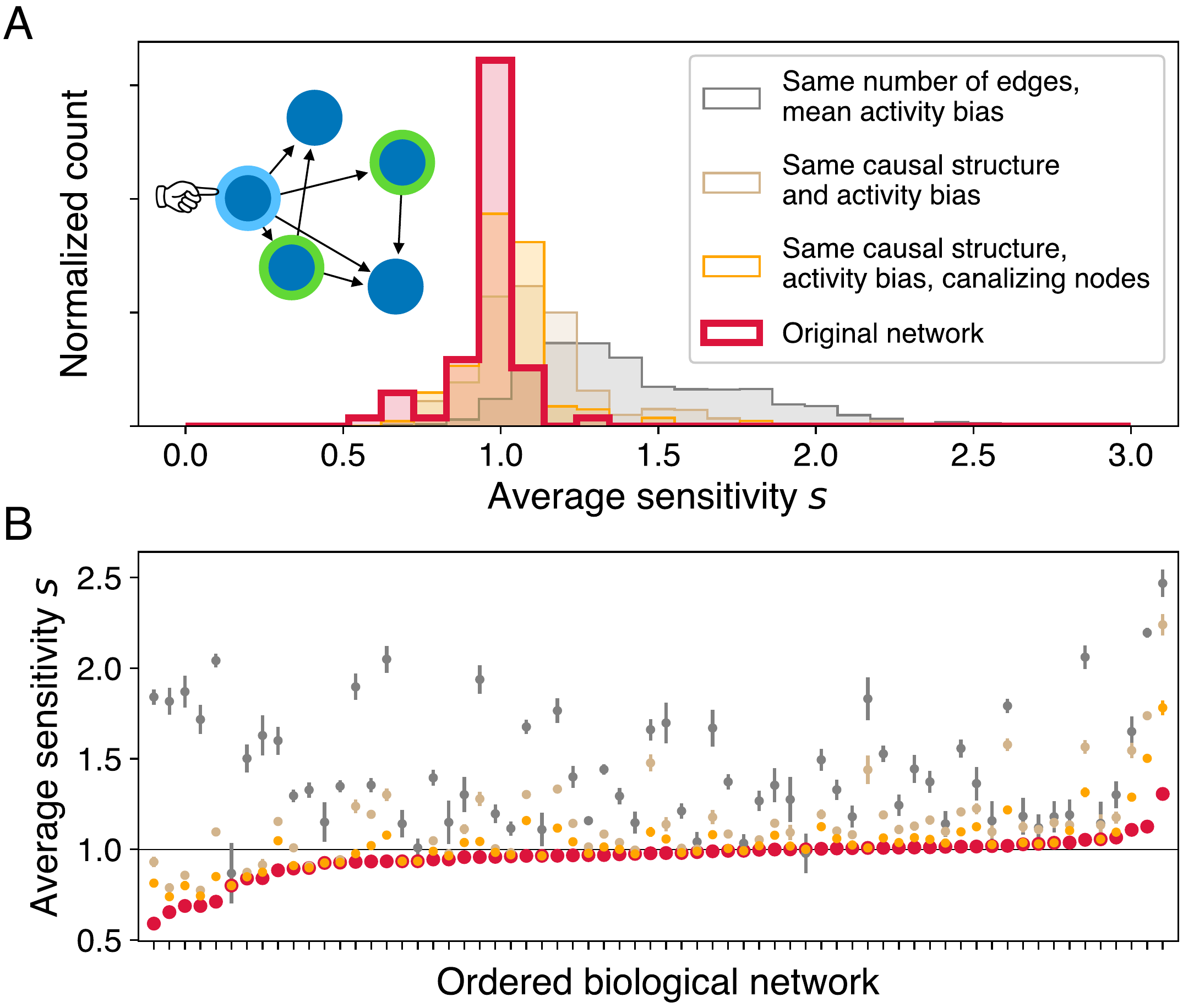}
\caption{ \textbf{Biological networks are close to critical sensitivity.} (A) 66 published Boolean network models of genetic regulation
(red) have sensitivity near the critical value of 1.
The schematic depicts the sensitivity measure, equal to
the average number of nodes whose states are changed
at timestep $t+1$ (green) when one node's state is
changed at timestep $t$ (light blue). 
Also shown are sensitivities of 
random ensembles preserving various aspects of the
original networks.  Preserving only the number of edges
and mean activity bias (gray) produces much more chaotic 
networks.  Preserving the causal structure and activity
bias of each node in the network (tan) produces 
sensitivity generally nearer to 1, and further
restricting to have the same number of canalizing 
functions (yellow) is closer still.  This indicates that
the specific structure and types of Boolean functions 
beyond average connectivity are important to criticality. 
(B) \textbf{Naive random Boolean network theory does not correctly predict average sensitivity for most biological network models.}  Plotting average sensitivities for each network and its random ensembles separately reveals that 
most networks have sensitivity that is significantly different
from that predicted by various random ensembles.  The mean
and standard deviation for each ensemble is shown 
for 100 samples from each ensemble.
\label{sensitivityCombined}%
}
\end{figure}

We find biological networks have stricter conditions on their causal structure and logic necessary to explain their criticality than merely constraining average properties such as $\langle K \rangle$ and $\langle p \rangle$. In particular, our results demonstrate biological networks are atypical within the ensemble of random networks with fixed $\langle K \rangle$ and $\langle p \rangle$ in three ways: the Boolean functions in each biological network typically 
(1) display covariance in $K$ and $p$, in that functions with
larger in-degree tend to have smaller $ p(1-p)$; 
(2) depend on all $K$ inputs (making connectivity coincide with causal structure); and
(3) are mostly canalizing.
Each of these constraints significantly affects the sensitivity, 
yet we nonetheless find that all biological networks we measure have
sensitivity near one (\figref{sensitivityCombined}). Our results indicate living networks are more distinguished by their criticality than  other network properties such as the degree distribution, edge density, or fraction of activating conditions. 









\begin{figure}
\centering
\includegraphics[scale=0.4]{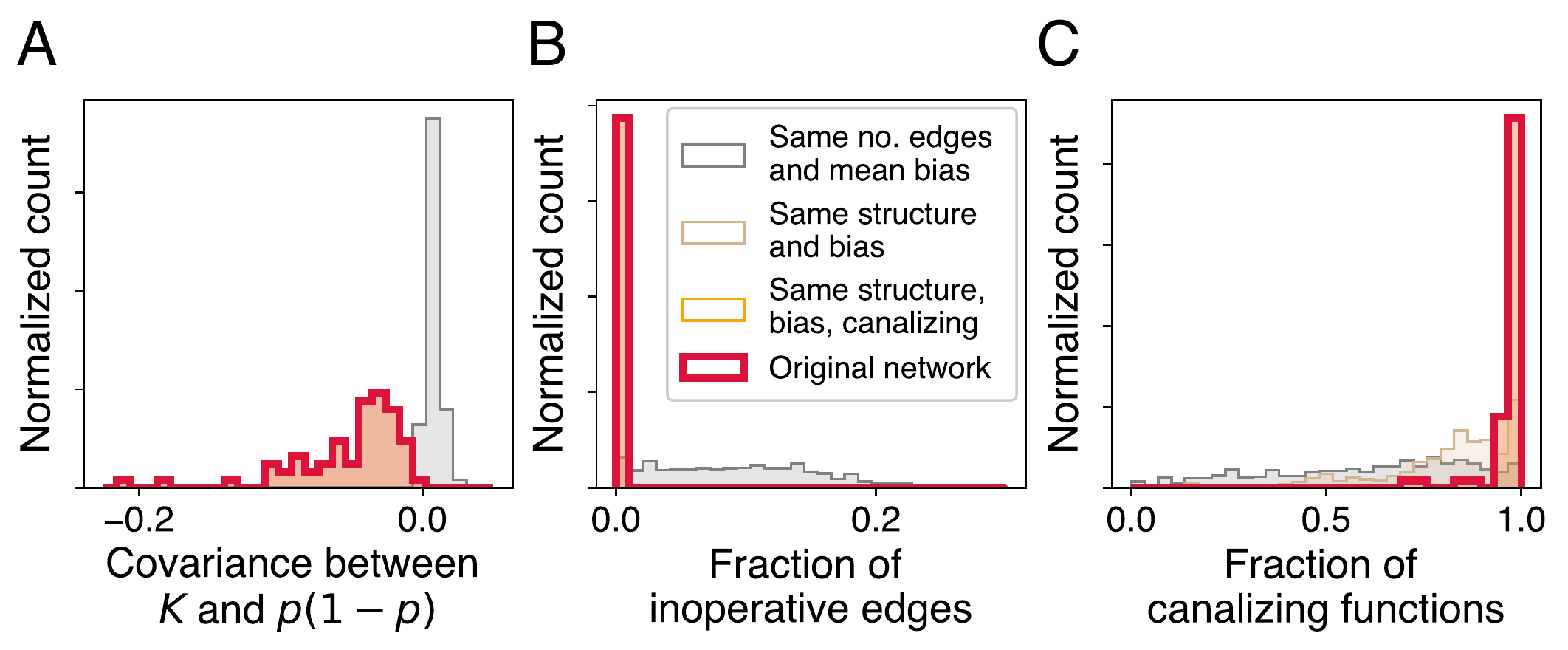}
\caption{ \textbf{Boolean models of biological networks are atypical in theoretical random ensembles.}
(A) The activity bias $p$ and in-degree $K$ covary in observed networks, such that the simple expected sensitivity in constant $p$, constant $K$ networks [\eqref{snaive}] is not valid.  
(B) Models of biological networks do not include 
inoperative edges in which the output does not depend on the input, whereas theoretical ensembles typically do.
(C) Compared to uniform sampling over Boolean functions, a much larger fraction of functions in biological networks are canalizing.
\label{atypicalAspects}%
}
\end{figure}

To determine the properties of gene regulatory networks most important for achieving their near-critical state, we compare analysis of the sensitivity of the biological networks to the same analysis performed on three different ensembles of  random networks. The random ensembles used in this study were designed to to successively isolate the properties of the real gene regulatory networks that drive their criticality, and are therefore constructed with reference to each of the 66 biological models (such that there are 66 different random ensembles, one for each gene regulatory network model, for each random network variant tested; see \figref{sensitivityCombined}B). In previous work, most ensembles of random Boolean networks have been defined such that the probability of a given node $i$ to 
be activated by a given condition, the activity bias $p_i$, is equal for all nodes.
The average sensitivity $s_i$ can be calculated 
for each node separately (such that $s = \sum_i s_i/N$),
and it is known that, when calculating sensitivity for each Boolean function with $K_i$ inputs and activity bias $p_i$, 
$\avg{s_i} = 2 K_i p_i (1 - p_i)$,
where the average is taken over possible Boolean functions \cite{ShmKau04}.
When naively assuming $p_i = p \, \forall \, i$, or more generally when
$K_i$ is not correlated with $p_i(1- p_i)$, 
the average sensitivity for
the network is simple:
\begin{equation}
\label{snaive}
s_\mathrm{naive} = 2 \avg{K} \avg{p (1 - p)},
\end{equation}
where each average is taken over nodes $i$.
But as shown in \figref{atypicalAspects}A we find that most networks in our database 
display anti-correlation between $K_i$ and $p_i(1-p_i)$,
meaning that a more accurate estimate of the average
sensitivity for real genetic circuits would require knowledge of the magnitude of this
covariance:
\begin{equation}
s_\mathrm{random} = 2 \avg{K} \avg{p (1 - p)} + 2 C,
\end{equation}
where $C$ is the covariance over nodes between
$K_i$ and $p_i(1-p_i)$.
Random networks that 
conserve $\avg{K}$ and $\avg{p}$ but do not 
conserve the network structure (gray color
in \figref{sensitivityCombined}) corresponding to networks with the same global structure (same number of edges and mean activity bias) should therefore be expected to have very different $s$ than the biological networks, as we indeed observe.

We can conclude random networks conserving the global structure and logic of biological networks ({\it e.g.} same $\langle p \rangle$ and $\langle K \rangle$) do not reproduce the critical behavior of real gene regulatory networks. We therefore next constructed random network ensembles that control for the local causal and logical properties of biological networks, not just their global ones. Specifically, we note network structure encapsulates the causal interactions of each node defined by its number of inputs and outputs. To test the role of this structure in driving critical behavior, we constructed a random ensemble of Boolean networks that conserves the local causal structure of each node, changing only the specific Boolean
functions implemented by each node. That is, we constructed random networks keeping the same causal inputs and outputs and the same activity bias $p_i$ for each node, but with randomized Boolean logic functions (labeled as ``same causal structure and activity bias'').  This ensemble explicitly 
retains the covariance between $K$ and $p(1-p)$. 
It also retains the fact that all connected inputs in the biological models are ``operative'' in the sense that
there is at least one state for which the output value 
of the node depends on each input; this is not necessarily 
true in randomized networks (\figref{atypicalAspects}B). We find even this very restricted ensemble
(tan color in \figref{sensitivityCombined}) 
is often distinctly more chaotic than the original biological networks.
This is due to a third way the random ensembles
are distinct from biology: 
Biological regulatory networks are hypothesized to display an overabundance of functions
that are canalizing \cite{HarSawWue97,HarSawWue02,KauPetSam04}, meaning that these functions have at least one input that can be fixed to a value that forces the output to a specific value regardless of the other inputs.  We find that
the biological networks indeed overwhelmingly consist
of canalizing functions (\figref{atypicalAspects}C). As has been argued before \cite{HarSawWue02,KauPetSam03,KauPetSam04,RamKesYli05,CorGatWan18}, canalizing functions tend to have smaller sensitivity, 
meaning that random ensembles that do not take this into
account will appear more chaotic than those that sample from
canalizing functions. We test the role of canalizing functions in driving criticality by constructing a third ensemble of random networks, which control for causal structure, activity bias, and canalizing nodes (yellow color in 
\figref{sensitivityCombined}). This ensemble most closely matches the criticality observed in the biological networks of the ensembles tested. 

In sum, we find that knowing only mean properties of biological networks (mean in-degree $\langle K \rangle$ and activity bias $\langle p \rangle$) is not enough to predict criticality.  Nor is knowing the exact network structure.  We must additionally include Boolean functions that depend on all their inputs and tend to be canalizing in order to model gene circuits with biological function.


Our results confirm an average sensitivity close to the critical value of one is sufficient to distinguish biological gene regulatory networks from random networks with similar global structure and logic. This suggests the most distinguishing features of biological networks are not their global connectivity patterns, such as degree distribution or edge density, or even the average bias of their logic operations. Instead criticality in gene regulatory networks is explained in terms of their local causal and logical structure, quantified in terms of the covariance of $K$ and $p(1-p)$ and a much higher frequency of canalizing functions in their implemented logic than naive models would predict. While critical sensitivity is a collective property of the interactions of many components, we find its explanation in gene regulatory networks relies on constraining the local causal and logical structure of individual nodes. This suggests that evolution is optimizing  the macroscale behavior of gene regulatory networks, as quantified by their criticality, by tuning the microscale interactions of individual genes.

Better understanding the relationship between $K$ and $p(1-p)$ for critical biological networks should inform evolutionary models and provide testable criteria for assessing criticality of gene regulatory networks in the lab. For example, in order for a genetic circuit to be critical, our results indicate genes that are regulated by many others must remain largely insensitive to many of their inputs. Our results confirm that neither network structure nor logic alone can predict the behavior of biological networks, and that knowing both 
is necessary to understand their behavior. In this sense, criticality in biological networks, which captures something of their collective properties, can be considered as an emergent property of logic and causal structure taken together. This has implications for our understanding of the physics of living processes where the connection between information processing (aggregate logic) and causation (aggregate connectivity) has yet to be fully explicated. 

This project/publication was made possible through support of a grant from Templeton World Charity Foundation. The opinions expressed in this publication are those of the author(s) and do not necessarily reflect the views of Templeton World Charity Foundation.

\bibliographystyle{unsrt}
\bibliography{grn}

\section*{Supplemental information}

\subsection*{Computational methods}

All analysis used for this work was carried out using a custom python package, Neet \cite{Neet5a45a92}, developed at Arizona State University.
Neet implements a collection of dynamical network types, a suite of network analyses including average sensitivity, and the randomization techniques described below.

\subsection*{Network Randomization}\label{subsec:randomizations}
We generated ensembles of random networks for each biological network by varying the base network subject to specific constraints.
Each constraint restricted the space of admissible Boolean networks, and the networks were selected with uniform probability over these subspaces.
All of the ensembles were constructed by first constraining some aspect of the biological network's topology and allowing all others to vary.
For this study, we considered two topological constraints: fixed mean degree and fixed causal structure. The fixed mean degree ensemble ensures only that each random network has the same number of total edges as the biological network, but otherwise the properties of individual nodes are not conserved. Random networks maintaining a fixed causal structure are topologically identical \footnote{This is a stronger constraint than simply graph isomorphism.} to the reference biological network and therefore share global properties such as mean degree, but also the local connectivity of each node, with the biological network. 

One subtlety to our approach is that, regardless of which topological randomization we considered, external nodes were always conserved.
In this work, an \textit{external node} is defined as any node in the original, biological network that has no incoming connections.
These nodes typically represent environmental parameters such as temperature, pH or other chemical compounds that are not influenced by the network.
As we wish to preserve the special roles of the external nodes, we forbid any incoming edges to these nodes when rewiring the networks during randomization, and do not include them when calculating the average sensitivity or the metrics in \figref{atypicalAspects}.

Once the topology of the network was constructed, a Boolean function was randomly, though not uniformly, selected for each node consistent with the topology of the network.
For a node with in-degree $K$, we select a function by generating $2^K$ Boolean values, one for each possible input to the function, with a bias $p$ toward activating conditions, so that a total of $2^Kp$ Boolean values activate the node. In cases where $2^Kp$ is not an integer, its floor and its ceiling are randomly selected according to their closeness to $2^Kp$, with probabilities $\lceil 2^Kp \rceil - 2^Kp$ and $2^Kp - \lfloor 2^Kp \rfloor$ respectively.
We considered two methods of selecting the bias $p$ for each node:
(1) When fixing the mean activity bias, we use the average bias over nodes in the biological network as the bias for each of the random network's nodes. (2) When fixing activity bias at the node level, each node in the random network retains the bias of the associated node in the biological network.

It has been noted in previous work \cite{KauPetSam03} that generating Boolean functions that are compatible with a given network topology can result in inoperative edges.
An \textit{inoperative edge} is one in which the source bears no logic influence to the target.
These inoperative edges are readily created when selecting truth tables uniformly at random, particularly when the in-degree of the function is small.
When attempting to randomly select a truth table for a Boolean function with two inputs, i.e. a node with in-degree two, there is a 37.5\% chance that at least one of the inputs will have no effect on the output of the function, e.g. Table \ref{table:truth-table-1}.
When we do not explicitly disallow inoperative edges, we find significant mismatches between the desired topological structure and the structure generated by the constructed Boolean functions (\figref{atypicalAspects}B).
In our randomization methods that constrain the 
causal structure, we disallowed inoperative edges.

\begin{table}[!ht]
\begin{tabular}{c|c|c}
\textbf{X} & \textbf{Y} & \textbf{Z} \\\hline
         0 &          0 &          1 \\\hline
         1 &          0 &          1 \\\hline
         0 &          1 &          0 \\\hline
         1 &          1 &          0
\end{tabular}
\caption{\label{table:truth-table-1} An example of an inoperative edge in a truth table for a node $Z$ which functionally depends on node $Y$ but not $X$. The value of $Z$ is independent of the value of $X$, and we therefore qualify the edge $X \rightarrow Z$ as inoperative. Situations such as this can readily arise when randomly sampling truth tables.}
\end{table}

Our final consideration was that of canalizing functions.
An input variable into a function with $K$ inputs is called \textit{canalizing} if there exists at least one value of the variable which fixes the output of the function.
In other words, the function produces the same value regardless of the value of the other $K-1$ variables.
A function is called \textit{canalizing} if it has at least one canalizing input variable.
We observe that the biological networks exhibit significant proportions of canalizing functions, \figref{atypicalAspects}(C).
In addition to all of the above constraints on our network ensembles, we also construct ensembles which preserve the canalizing nodes of the original biological network.
Particularly, if a node in the original biological network has a canalizing function, the associated node in the randomized network is also canalizing though not necessarily with the same canalizing Boolean function.

\subsection*{Cell Collective networks}

\begin{figure}
\centering
\includegraphics[scale=0.4]{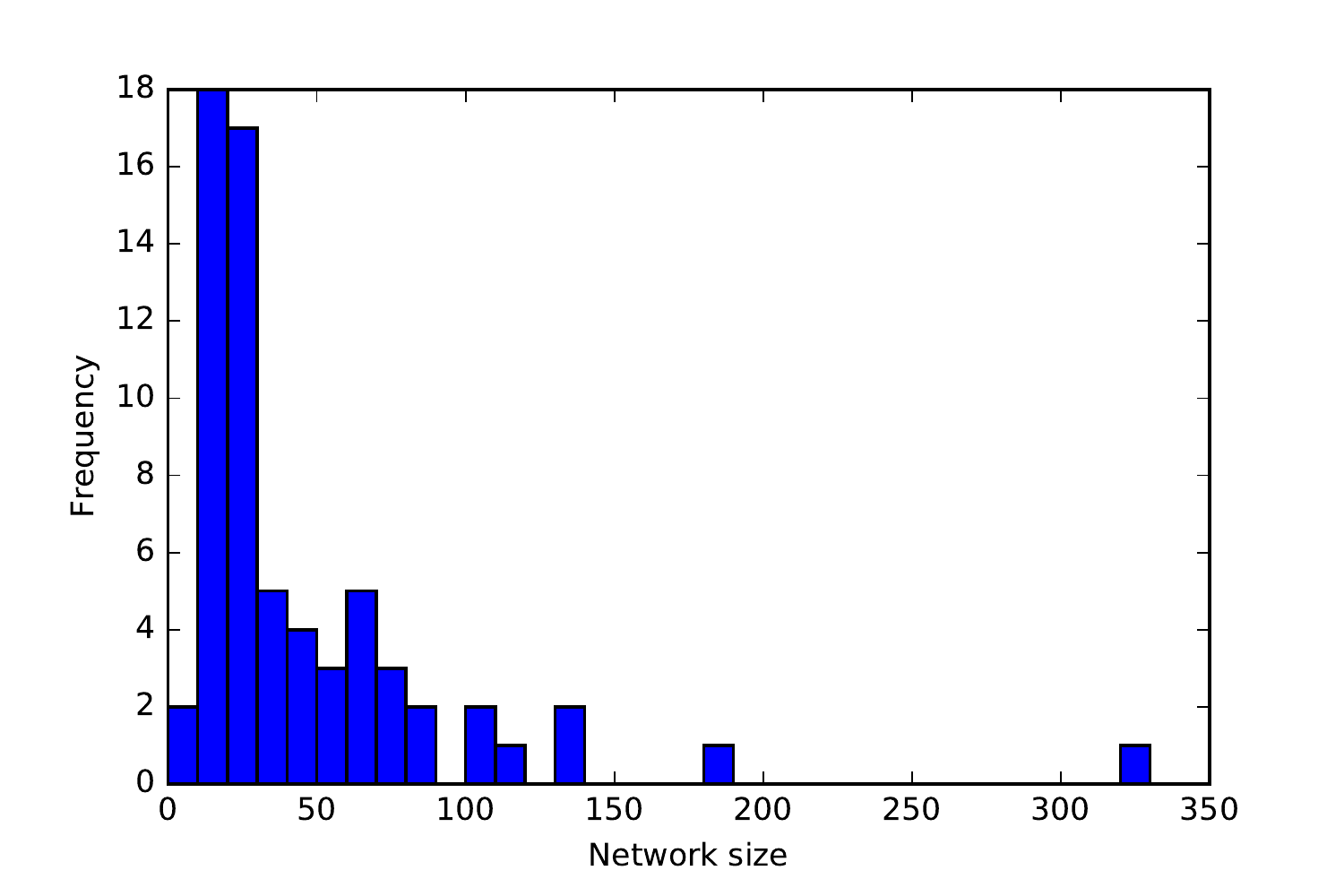}
\caption{ \textbf{Distribution of model network sizes.}
\label{networkSizes}%
}
\end{figure}

\begin{figure}
\centering
\includegraphics[scale=0.4]{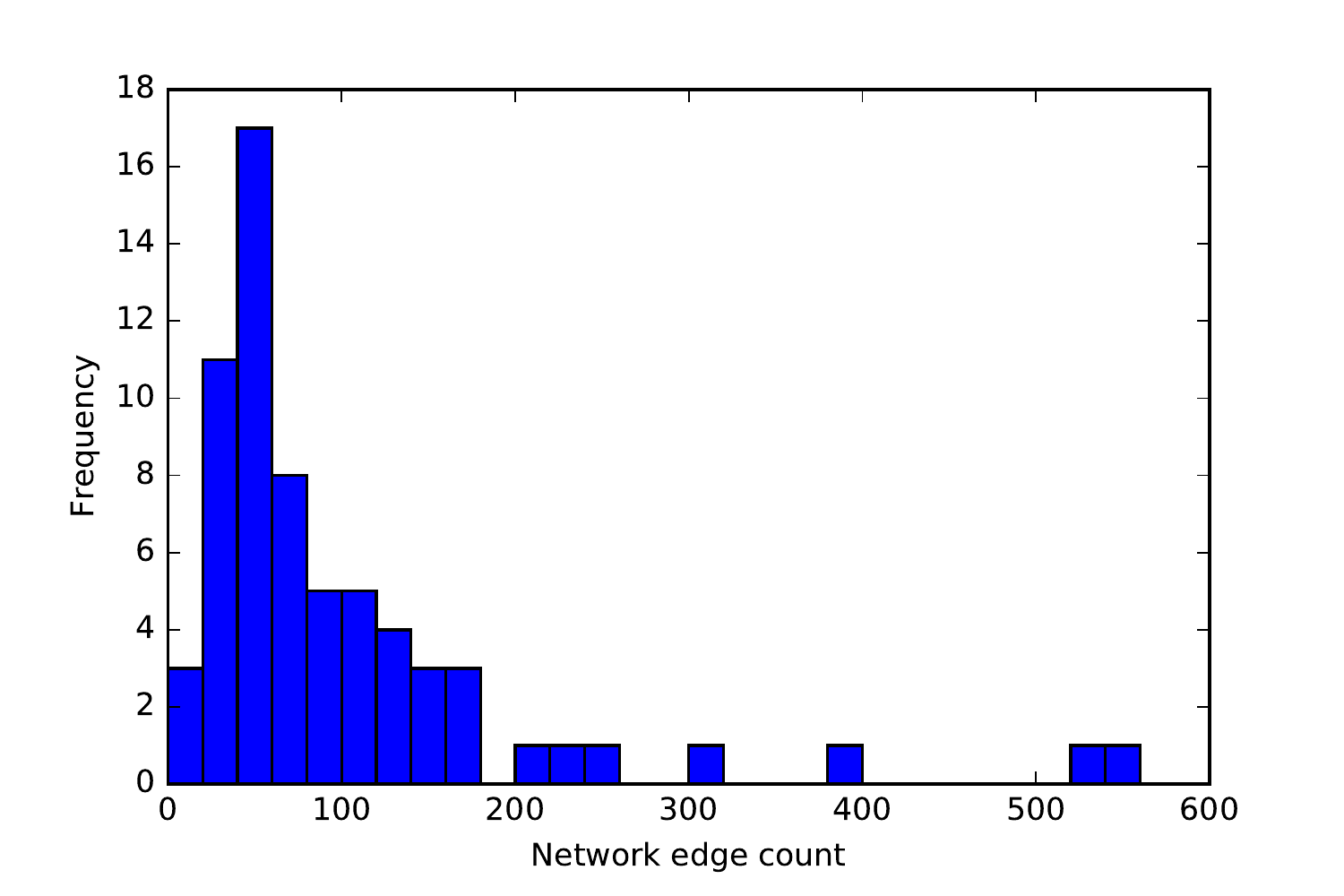}
\caption{ \textbf{Distribution of number of interactions in model networks.}
\label{networkEdges}%
}
\end{figure}

Boolean models were obtained from the Cell Collective database \cite{cellcollective} as logical expression and truth tables.
The Cell Collective is an online platform to build, analyze, and share biological network models \cite{helikar2012cell}.
While the Cell Collective provides tools to visualize and simulate dynamic biological networks, we use it only as a database of Boolean models.  
The biological processes represented include (but are not limited to) organism immune responses, pharmacodynamics, cancer progression (including breast, prostate, and blood), HIV progression, tumor development, cortical and cardiac development, cholesterol synthesis, viral-host interactions, plant stomatal function, DNA damage and repair, human sex determination, and various growth factors.
The database contains a mixture of networks which obey synchronous and asynchronous update rules, which we do not discriminate in our analyses.
We analyze 66 networks, all without inoperative edges, ranging in size $N$ from 5 nodes to 321 nodes.  Nodes typically represent genes and proteins, and occasionally represent higher-level biological states. Over half the networks are smaller than 27 nodes, and over 90\% of networks are smaller than 102 nodes (\figref{networkSizes}). The number of interactions per network scales from 15 to 551. Half of the 66 networks have less than 66 edges, and over 90\% of networks have less than 201 edges (\figref{networkEdges}).  

\begin{figure*}
\centering
\includegraphics[scale=0.4]{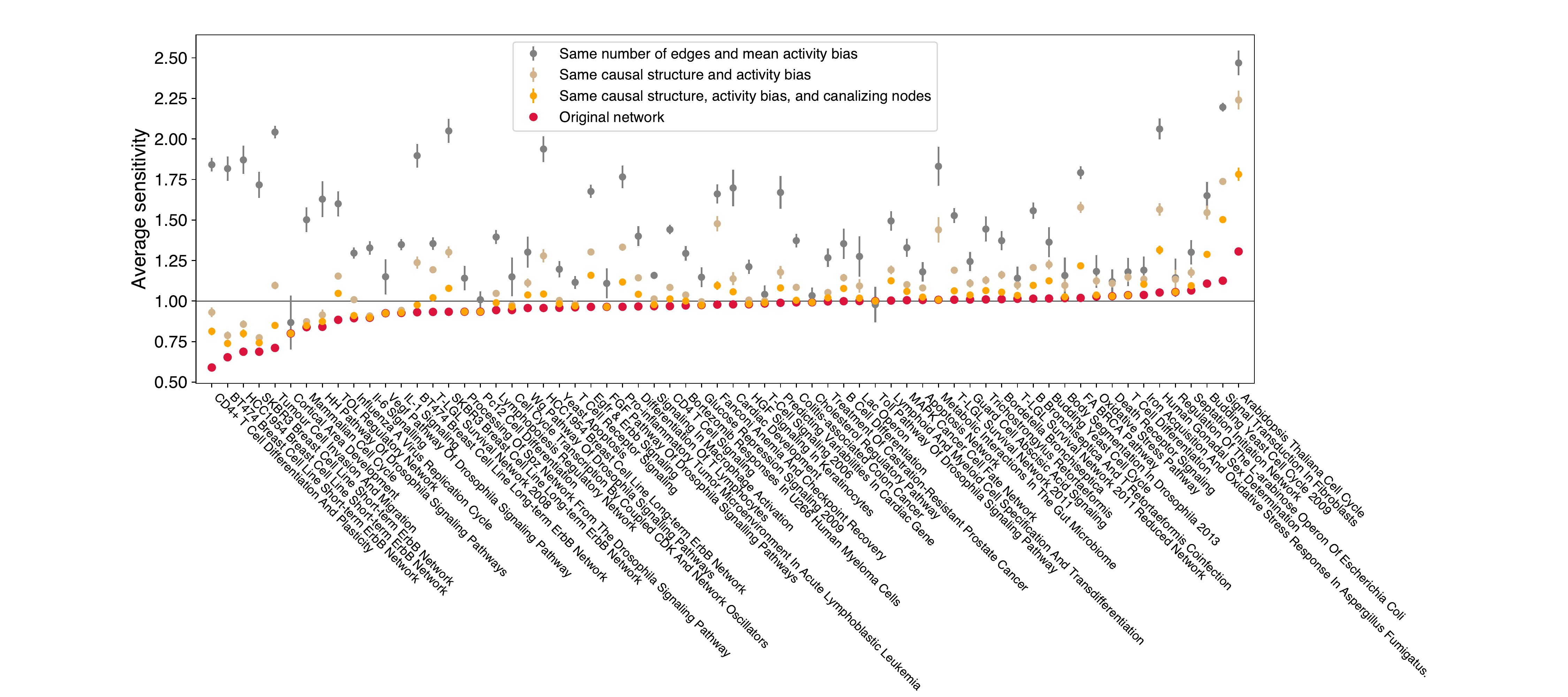}
\caption{ \textbf{\figref{sensitivityCombined}B including the names of the models.}%
\label{sensitivityWithNames}
}
\end{figure*}

All models available in the Cell Collective database at the time of retrieval are included in this study, except for three models that each include one function that does not depend on any of its inputs.
In \figref{sensitivityWithNames}, we replot \figref{sensitivityCombined}B with the corresponding 
name of each Cell Collective model used in the study.

\subsection*{Sensitivity as a function of in-degree}

\begin{figure*}
\centering
\includegraphics[scale=0.5]{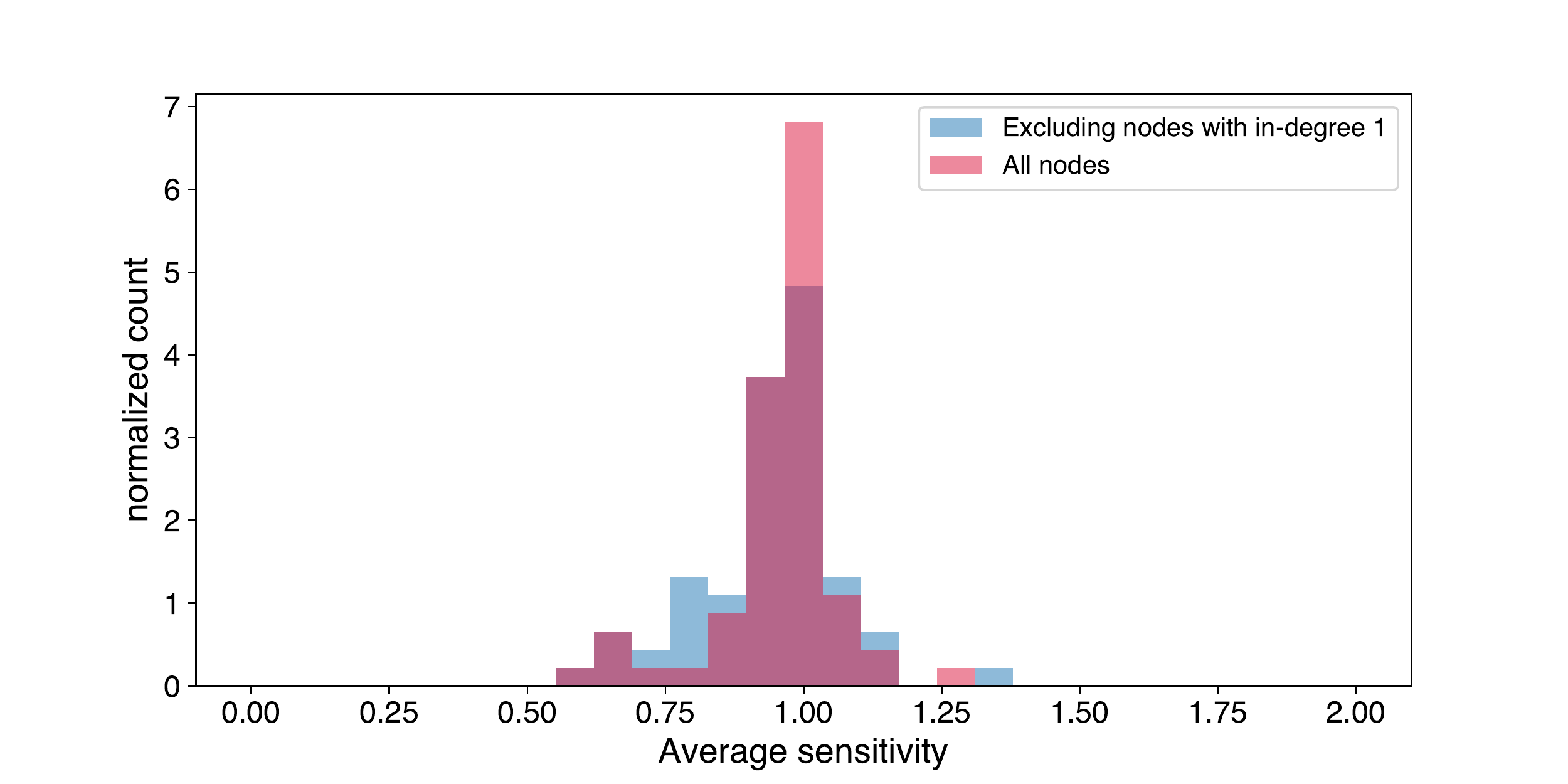}
\caption{ \textbf{Sensitivity excluding nodes with in-degree 1.}
\label{sensitivityNoOne}%
}
\end{figure*}

Of the
2,596 total non-external nodes in our dataset, 1,054 (about $40 \%$) have in-degree $K_i=1$.  Because we consider only
operative edges, nodes with $K_i=1$ always have sensitivity
$s_i = 1$.  To determine whether this biases our result
for average sensitivity over all nodes, we also compare
a calculation that excludes all nodes with $K_i = 1$
in \figref{sensitivityNoOne}.  Even excluding these
nodes produces a similar peak at $s=1$.

\begin{figure}
\centering
\includegraphics[scale=0.5]{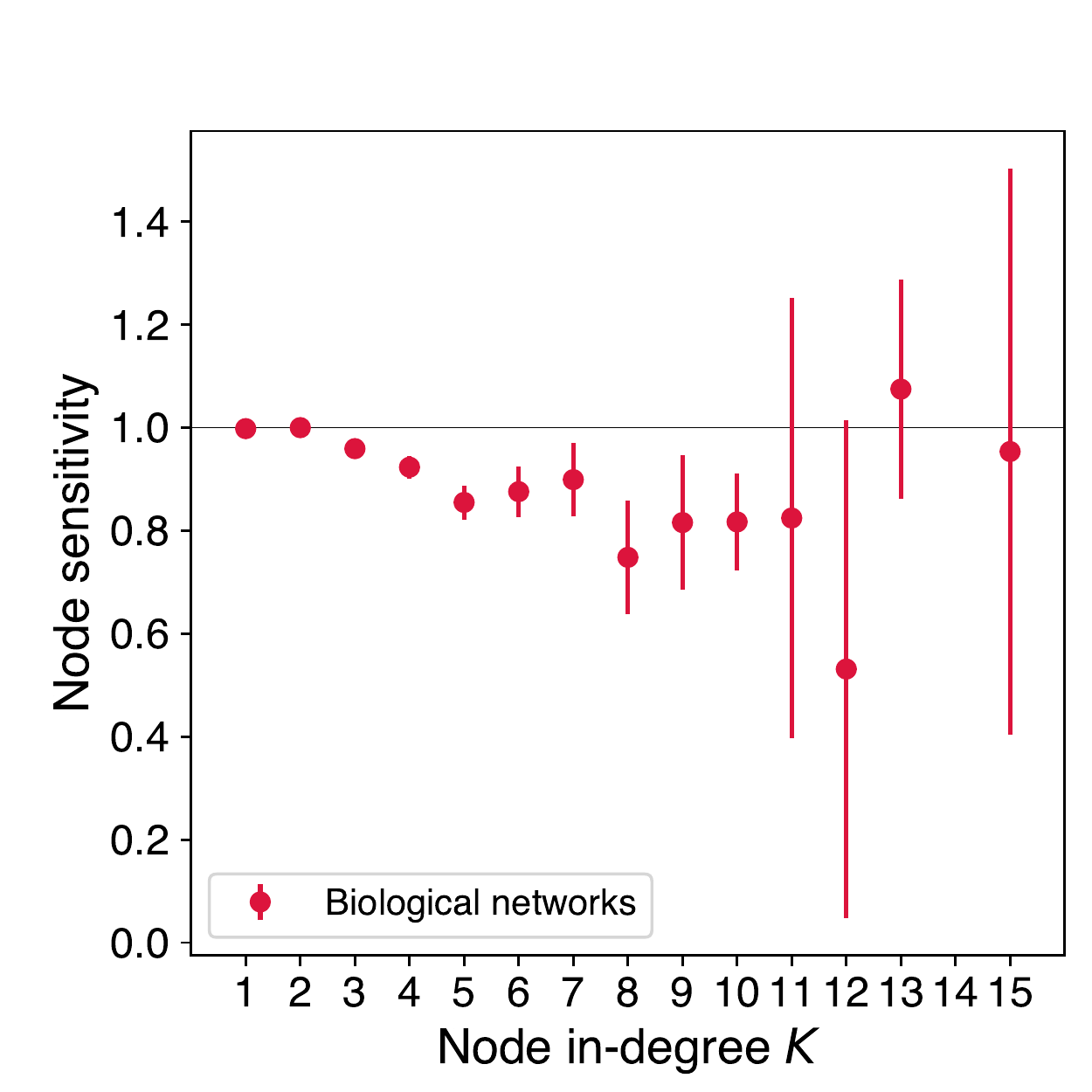}
\caption{ \textbf{Node sensitivity versus in-degree for biological networks.}  Error bars indicate standard error of the mean.
\label{sensitivityVsDegree}%
}
\end{figure}

We also plot the mean sensitivity of individual nodes as
a function of in-degree in \figref{sensitivityVsDegree}.

\subsection*{An equivalent definition of sensitivity}

For a given state $\vec x$, the number of nodes whose states
change when flipping $x_j$ is $\sum_i \frac{\partial f_i(\vec x)}{\partial x_j}$, which when averaged over $j$ and $\vec x$
gives an average Hamming distance of 
\begin{equation}
\frac{1}{N} \sum_j^N \avg{ \sum_i^N \frac{\partial f_i(\vec x)}{\partial x_j} }_{\vec x} = s
\end{equation}
after swapping the order of the sums over $i$ and $j$.

\end{document}